# Pairing anti-halo effect

K. Bennaceur,[1,2] J. Dobaczewski,[3] M. Płoszajczak[2]
[1] *Centre d'Etudes de Bruyères-le-Châtel, BP 12, F-91680 Bruyères-le-Châtel, France*
[2] *Grand Accélérateur National d'Ions Lourds (GANIL), CEA/DSM – CNRS/IN2P3, BP 5027, F-14076 Caen Cedex 5, France*
[3] *Institute of Theoretical Physics, Warsaw University, Hoża 69, PL-00681, Warsaw, Poland*

## Abstract

We discuss pairing correlations in weakly bound neutron rich nuclei, by using the coordinate-space Hartree-Fock-Bogolyubov approach which allows to take properly into account the coupling to particle continuum. We show that the additional pairing binding energy acts against a development of an infinite rms radius, even in situations when an $\ell$=0 single-particle orbital becomes unbound.

An essential element in understanding the physics of weakly bound fermion systems, is the residual coupling between bound states and continuum. With decreasing particle separation energy, the zone of correlated states around the Fermi surface must incorporate the phase space of particle continuum. One of the most interesting discoveries in loosely bound nuclei is the neutron halo [1,2]. In this low density, high-isospin state of matter, the two-particle correlations, like the pairing modified by the effects of virtual scattering to the continuum, are expected to play a crucial role.

Nuclear pairing correlations are important throughout the periodic table, providing an extra binding for ordinary even-even and odd nuclei, but they become essential for loosely bound nuclei, because without them the systems in question cease to be bound. For their description, two main elements have to be combined. Firstly, the residual interaction which is expected to be different in the nuclear exterior and in the nuclear interior, should include the density dependence in the self-consistent framework. Secondly, the residual coupling between bound and continuum states must be treated consistently. Both elements are present in the Hartree-Fock-Bogolyubov (HFB) coordinate-space, mean-field approach using the Skyrme energy density functional [3].

The asymptotic behavior of halo nuclei, was discussed both for 2-body [4,5] and 3-body systems [6,7]. For 2-body systems, the root-mean-square radius (rms) of a low-angular-momentum ($\ell$=0,1) neutron orbital in an average potential diverges when its energy approaches the continuum. Rigorous results for 3-body systems have been obtained neglecting the antisymmetrization. The strong divergencies seen in 2-body systems are now seldom. When all 2-body subsystems of the 3-body system are unbound, the rms radius diverges at most logarithmically with vanishing binding energy [6]. When one subsystem has a bound state with energy close to the total energy of the 3-body system, the system behaves as one particle in the common field of the 2-body state [7]. Finally, when at least two subsystems have bound states, the ground state (g.s.) rms radius always converges [7].

The pairing near the neutron drip line was studied first in a schematic 3-body model of two interacting nucleons together with a structureless core [8]. The exponential decay of both the single-particle (s.p.) density and the 2-particle density for valence neutrons was found to be governed rather by the 2-particle energy of the correlated pair than by the s.p. binding energy. This analysis is insufficient because just near the drip lines the particle-hole (p-h) and particle-particle (p-p) channels of the effective interaction are strongly coupled, and the non-self-consistent treatment of the pairing interaction based on computing pairing correlations after determining the s.p. basis is unjustified.

In light nuclei, many-body correlations may essentially modify the mean-field picture [9,10]. For that, different cluster models have been applied [11,12], and the 3-body dynamics has been investigated [11,13–15]. Much progress has been made in the treatment of the Pauli exclusion principle [16], which is needed to remove spurious components of the halo wave function even at small densities. The whole problem is subtle because the proper antisymmetrization of the wave function easily spoils many advantages of the cluster description. It was found that the 3-body asymptotics of the cluster wave function is given by the exponential fall-off with a decay constant $\kappa{=}\sqrt{2mE_{(3)}}/\hbar$, where $E_{(3)}$ is the binding energy of the system relative to the 3-body break-up threshold. If a bound 2-body subsystem is present, the asymptotics is given by a binary channel asymptotics [17].

In heavier nuclei, the concept of self-consistent mean-field is better justified. Moreover,

the one-body densities are always governed by the so-called natural orbitals (the eigenstates of the exact one-body density matrix), and the mean-field approximation can be understood as a method to analyze properties of the natural orbitals. Strong pairing correlations do not allow for the decoupling of weakly bound neutrons from the rest of the system [18,19]. It is therefore a crucial question whether the neutron halo phenomenon can happen close to the neutron drip-line in heavy nuclei.

In the case of no pairing, irrespective of whether the number of neutrons is even or odd, the g.s. neutron mean-field asymptotic density equals $\rho(r){\sim}\exp(-2\kappa r)/r^2$ for $\kappa{=}\sqrt{-2m\varepsilon}/\hbar$ ($\varepsilon$ is the s.p. energy of the least bound $\ell{=}0$ neutron), and has a long tail with the rms radius $r_{\rm rms}{\sim}1/\kappa{\to}\infty$ for $\varepsilon{\to}0$.

In the presence of pairing, we have to separately consider the even-$N$ and odd-$N$ isotopes. For $N$ even, the asymptotics of the HFB neutron density is given by the binding energy relative to the 2-body break-up threshold $E_{(2)}$. Therefore, the decay constant becomes [20,21]: $\kappa{=}\sqrt{2mE_{(2)}}/\hbar$ for $E_{(2)}{=}\min[E_\mu]-\lambda$, where $\min[E_\mu]$ is the lowest *discrete* ($E_\mu{<}{-}\lambda$) quasi-particle (q.p.) energy, and $\lambda$ is the chemical potential. The HFB q.p. energies $E_\mu$ are fairly close to the canonical q.p. energies [21]:

$$E_\mu \simeq E_\mu^{\rm can} \equiv \sqrt{(\epsilon_\mu^{\rm can}-\lambda)^2+(\Delta_\mu^{\rm can})^2}, \tag{1}$$

where $\epsilon_\mu^{\rm can}$ and $\Delta_\mu^{\rm can}$ are the diagonal matrix elements of the p-h and p-p mean fields in the canonical basis. Hence, a canonical state at the Fermi surface ($\epsilon_\mu^{\rm can}{\simeq}\lambda$) gives the asymptotic density for $E_{(2)}{=}\Delta_\mu^{\rm can}{-}\epsilon_\mu^{\rm can}$, which is larger by $\Delta_\mu^{\rm can}$ than that in the no-pairing approximation. Therefore, in even-$N$ nuclei, the paired densities decrease faster than the unpaired densities, which is the effect called here the 'pairing anti-halo effect'.

In odd-$N$ nuclei, the neutron densities are given by one-q.p. states [22], which gives the asymptotic decay corresponding to $E_{(2)}=-\min[E_\mu]-\lambda$. This threshold energy goes to zero when the minus Fermi energy approaches the pairing gap of a state at the Fermi surface. Therefore, the pairing correlations do not prevent the neutron radii of odd-$N$ nuclei from diverging in the limit of $\lambda{\simeq}\,\epsilon_\mu^{\rm can}{\to}{-}\Delta_\mu^{\rm can}$.

Below we give a numerical example illustrating the mechanism of the pairing anti-halo effect in even-$N$ nuclei, which takes the most dramatic form for $\epsilon_\mu^{\rm can}{\simeq}\lambda{\to}0$, because then the no-pairing tail of the neutron density gives a divergent rms radius, while that for the paired density stays finite. Before that we must, however, consider that in weakly bound nuclei the situation is more complicated, because for $\lambda{\to}0$ there can be no *discrete* HFB quasiparticle state at all, and we have to take into account the HFB continuum explicitly. To this end we express the asymptotic density, coming from the q.p. continuum, in the form of the integral:

$$\rho(r) \sim \int_{-\lambda}^{\infty} dE \frac{A^2(E)}{r^2}\exp(-2\kappa(E)r), \tag{2}$$

where $\kappa(E){=}\sqrt{2m(E-\lambda)}/\hbar$, and $A(E)$ is the amplitude of the exponential tail corresponding to the quasiparticle having the q.p. energy $E$. It is obvious that at $r{\to}\infty$ the upper limit of the integral (2) is not important, and the asymptotic behavior of the density is determined by amplitudes $A(E)$ near the threshold of $E{=}{-}\lambda$.

These amplitudes are determined by the pairing interaction and cannot be inferred from any simple geometric considerations. We can expect that $A(E){\to}0$ for $\kappa(E){\to}0$, because in this limit, the lower HFB q.p. wave functions are completely delocalized [18], and thus become decoupled from the pairing field inside the nucleus. It is instructive to realize that a power-law dependence of $A^2(E){\sim}(E{-}\lambda)^\alpha$ gives for $\lambda{=}0$ the density vanishing according to a power law too: $\rho(r){\sim}r^{-4-2\alpha}$. At any small non-zero value of $\lambda$, the latter power law holds up to $r{\simeq}1/\kappa(-\lambda)$, but then turns over to the standard exponential dependence of $\exp(-2\kappa(-\lambda)r)$.

The two-neutron drip line ($\lambda{\simeq}0$) in heavier nuclei will not probably be reached before soon, however, the experimental data for neutron halos exist in light nuclei [23]. Fig. 1(a) compares the calculated one-neutron separation energies, obtained using the spherical HFB code of Ref. [20], with the experimental and systematics values in heavy carbon isotopes [24]. The Skyrme SLy4 interaction [25] in the p-h channel together with the zero-range, density-dependent (surface) pairing p-p interaction, as described in Refs. [21,26], has been used in these calculations. The overall agreement is encouraging, if one considers that the spherical symmetry is assumed and, moreover, correlations beyond the mean-field are very important in these light systems.

The HFB code has been improved by employing the analytical wave functions in the asymptotic region, which allowed for solving the HFB equations in a very large box, and thus obtaining an arbitrarily dense discretized-continuum spectrum. The box of $R_{\mathrm{box}}{=}125\,\mathrm{fm}$ has been used in the present study, which gives the lowest discretized-continuum HF state at $\varepsilon{\simeq}(\pi^2\hbar^2)/(2mR^2_{\mathrm{box}}){\simeq}0.013\,\mathrm{MeV}$. Therefore, we may safely treat the mean-field continuum phenomena for the Fermi energies $-\lambda$ at least several times larger than this limiting value of 0.013 MeV. We have also performed test calculations in the box of $R_{\mathrm{box}}{=}400\,\mathrm{fm}$, for which this limiting value is an order of magnitude smaller, and we found no modifications of conclusions reached in the present study.

In Fig. 1(b) the spherical self-consistent Hartree-Fock (HF) neutron s.p. energies are shown for $Z{=}6$ as functions of the neutron number $N$. The carbon isotopes are special in that the least bound spherical neutron orbital at the neutron drip line has $\ell{=}0$, i.e., infinite rms radius should appear in the no-pairing scenario, had this orbital approached the zero binding energy. Similarly, an infinite rms radius would be given by an unpaired weakly bound *deformed* orbital, had this orbital contained even a small $2s_{1/2}$ component.

In self-consistent calculations within a given effective interaction, it is extremely difficult to change the relative positions of the s.p. orbitals without spoiling other global characteristics of the weakly bound nuclei. Therefore, in order to study dependence of the results on the $2s_{1/2}$ binding energy, we have performed the HFB calculations for fixed (i.e., non-self-consistent) s.p. wave functions of the Pöschl-Teller-Ginocchio (PTG) potential (see the description and references given in Ref. [18]). Parameters of the PTG potential have been fixed at $\Lambda{=}5$, $s{=}0.085\,\mathrm{fm}^{-1}$. The depth parameter of the PTG potential for the $s_{1/2}$, $p_{1/2}$, $p_{3/2}$, $d_{3/2}$, and $d_{5/2}$ channels is $\nu_{\ell j}{=}3.251$, 2.814, 3.126, 2.747, and 3.176, respectively. At $N{=}16$, these values reproduce the HFB+SLy4 neutron spectrum of Fig. 1(b) exactly. Using the PTG potential we may now move the $2s_{1/2}$ orbital up or down, by changing the value of the depth parameter $\nu_{s1/2}$.

Figure 2(a) shows the s.p. energy of the $2s_{1/2}$ orbital as function of the potential depth $\nu_{s1/2}$. With decreasing $\nu_{s1/2}$, the $2s_{1/2}$ orbital approaches the zero binding energy at $\nu_{s1/2}{=}3$ (full circles), then turns into a virtual state (dashed line), and finally reappears at $\nu_{s1/2}{=}2.857$ as a very broad resonance (open circles). The rms radius of this orbital becomes infinite at $\varepsilon$=0, and hence the neutron PTG rms radius shown in Fig. 2(b) for $N$=16 (no pairing) becomes infinite too. On the other hand, the HFB rms radius which is also shown in Fig. 2(b), is completely unaffected by the transition of the $2s_{1/2}$ level through the point of no s.p. binding. This constitutes a vivid illustration of the pairing anti-halo effect. Detailed behavior of the rms radius can be traced back to the behavior of the $2s_{1/2}$ q.p. energies (1) induced by the canonical s.p. energies $\epsilon_\mu^{\rm can}$ and pairing gaps $\Delta_\mu^{\rm can}$ shown in Figs. 2(a) and (c), respectively.

The HFB calculations have to be performed not at a fixed particle number of $N$=16, but at a fixed Fermi energy of $\lambda$=−0.056 MeV. This value of $\lambda$ corresponds to $N$=16 at $\nu_{s1/2}$=3.251. At smaller values of $\nu_{s1/2}$, the $2s_{1/2}$ orbital depopulates, and the particle number cannot be kept constant for any negative value of $\lambda$. Values of the average neutron number $\langle N\rangle$ at $\lambda$=−0.056 MeV (see Fig. 2(c)), change from 16 to about 14.7 in the studied interval of $\nu_{s1/2}$. Nevertheless, the canonical-basis occupation factor $v^2_{\rm can}$ [21] of the $2s_{1/2}$ canonical state (open squares in Fig. 2(c)), is far from zero and decreases only to about 0.16 at the smallest value of $\nu_{s1/2}$=2.778. We have also performed calculations for fixed $N$=14 (this value can be kept constant with the varying position of the $2s_{1/2}$ orbital), and we obtained exactly the same qualitative results as for fixed $\lambda$ and a variable neutron number. The occupation of the $2s_{1/2}$ canonical state for $\lambda\simeq0$ is entirely due to virtual pair excitations.

In order to visualize the origins of the pairing anti-halo effect, in Fig. 3 are shown the contributions, $N(E)$ and $r^2(E)$, of the $\ell$=0 continuum HFB q.p. states to the particle number and radius squared, respectively. They are defined as the corresponding space integrals of the lower components [21] of the q.p. wave functions squared, $|\varphi_2(E,r)|^2$. When the $2s_{1/2}$ orbital is well bound (see panels (a) and (d)), the largest contributions to the particle number $\langle N\rangle$=$\int_{-\lambda}^{\infty}dEN(E)$, and to the radius squared $\langle r^2\rangle$=$\int_{-\lambda}^{\infty}dEr^2(E)$, come from narrow peaks located near the corresponding s.p. energies. However, when the $2s_{1/2}$ orbital is pushed across $\varepsilon$=0, and up into the continuum, the peaks in $N(E)$ and $r^2(E)$ do not at all approach $E$=0, but stay at a fairly large values of $E\simeq\Delta_\mu^{\rm can}\simeq2$ MeV. In this sense, the large pairing gaps prevent the largest contributions to the rms radius from entering the region of divergencies at $E\simeq0$, where the lower components of the HFB q.p. wave functions may develop long tails.

The sequence of curves presented in Figs. 3(d)–(f) is particularly instructive, because it shows that the $2s_{1/2}$ orbital gives either large or small contribution to the rms radius, depending on whether it is bound or resonant, but this contribution is always located around $E\simeq2$ MeV and simply disappears with $2s_{1/2}$ being pushed up. On the other hand, the low-energy continuum contributions, described by the asymptotic density matrices given in Eq. (2), are almost independent of the $2s_{1/2}$ position, and give well-pronounced, low-energy contributions to the rms radii. However, at any small but non-zero value of $\lambda$, the curve $r^2(E)$ bends over at small $E$, and goes to zero, thus giving a finite, non-diverging values of the HFB rms radii.

In conclusion, we have shown that pairing correlations significantly modify the asymptotic properties of one-body neutron densities $\rho(r)$ in weakly bound neutron rich nuclei. The neutron rms radii in even-$N$ nuclei are predominantly determined by two types of quasiparticles: (i) those near the threshold of $E$=−$\lambda$, and (ii) those corresponding to the $\ell$=0 s.p. orbitals. We have checked by numerical calculations that: (i) the density of the former quasiparticle states vanishes at $E$=−$\lambda$, which is due to a delocalization of the quasiparticle wave functions, and (ii) the latter quasiparticles have energies $E$ larger than the correspond-ing pairing gaps, and thus cannot give diverging rms radii. As a result, and contrary to the no-pairing approximation, the rms neutron radii in even-$N$ nuclei never diverge, even in the limit of $\lambda$=0.

In this work, we have analyzed the HFB asymptotics corresponding to the one-neutron densities. The 3-body asymptotics of two-neutron densities cannot be consistently described in the HFB. One should notice, however, that the break-up threshold for two-neutron emission is expected to be higher than the two-neutron separation energy $S_{2N}{\equiv}E^{(\text{even})}(N{-}2,Z){-}E^{(\text{even})}(N,Z)$ by the amount of the 2-particle energy of the correlated neutron pair. Hence, one expects that the pairing correlations will also act against the two-neutron halo effect in heavy even-$N$ systems, by preventing the neutron radii from rapid growth in the limit of $S_{2N}{\to}0$.

## ACKNOWLEDGMENTS

We would like to thank J.-F. Berger for useful discussions and W. Nazarewicz for a critical reading of the manuscript and many comments. This research was completed at the ECT* in Trento, Italy, and was supported in part by the Polish Committee for Scientific Research (KBN) under Contract No. 2 P03B 040 14 and by the French-Polish integrated actions program POLONIUM.

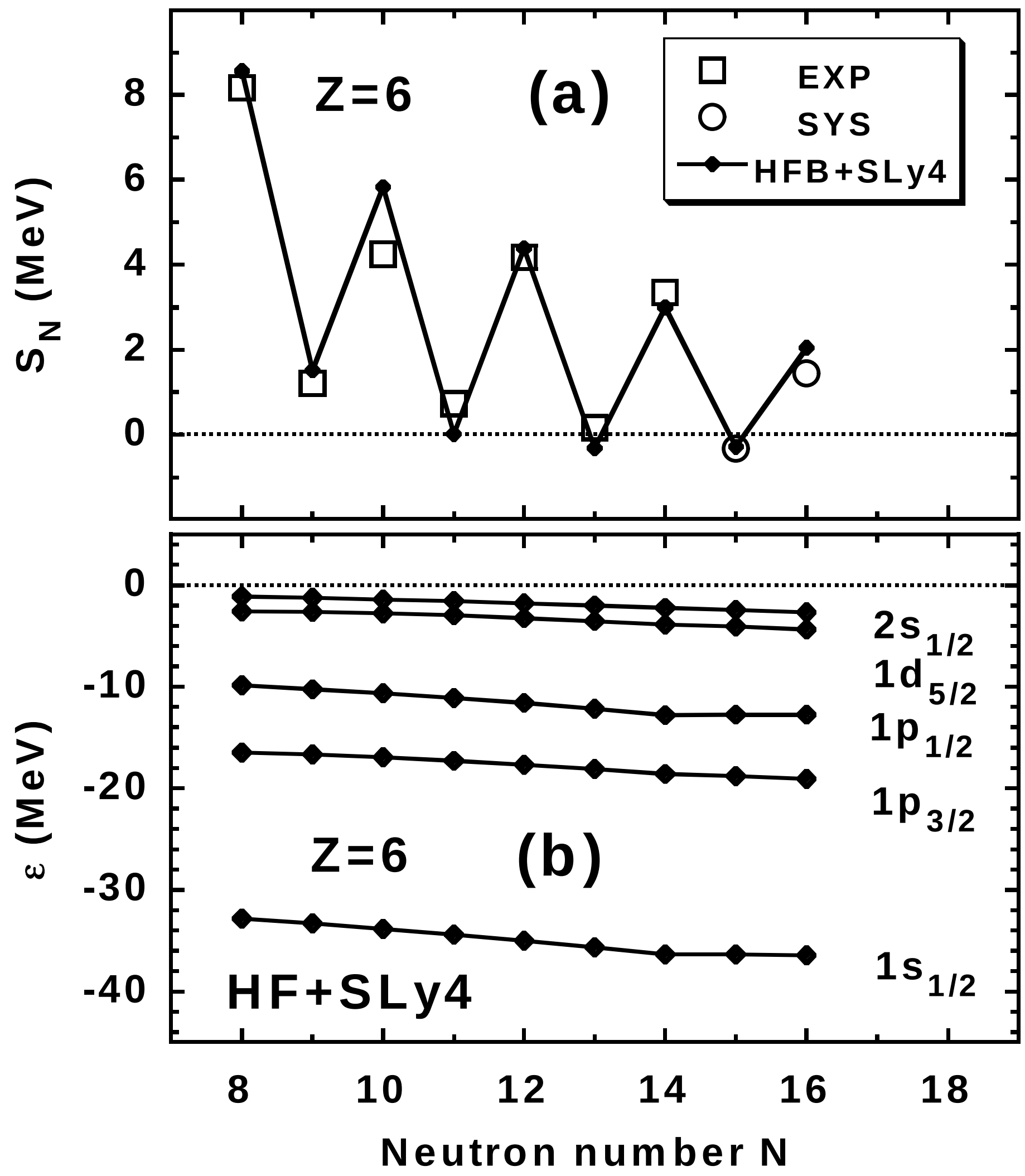

FIG. 1. One-neutron separation energies $S_N$ (a) and the spherical s.p. energies (b) in heavy carbon isotopes. Spherical self-consistent HFB and HF calculations have been performed for the Skyrme SLy4 interaction [25]. Experimental (EXP) and systematics (SYS) separation energies are taken from Ref. [24].

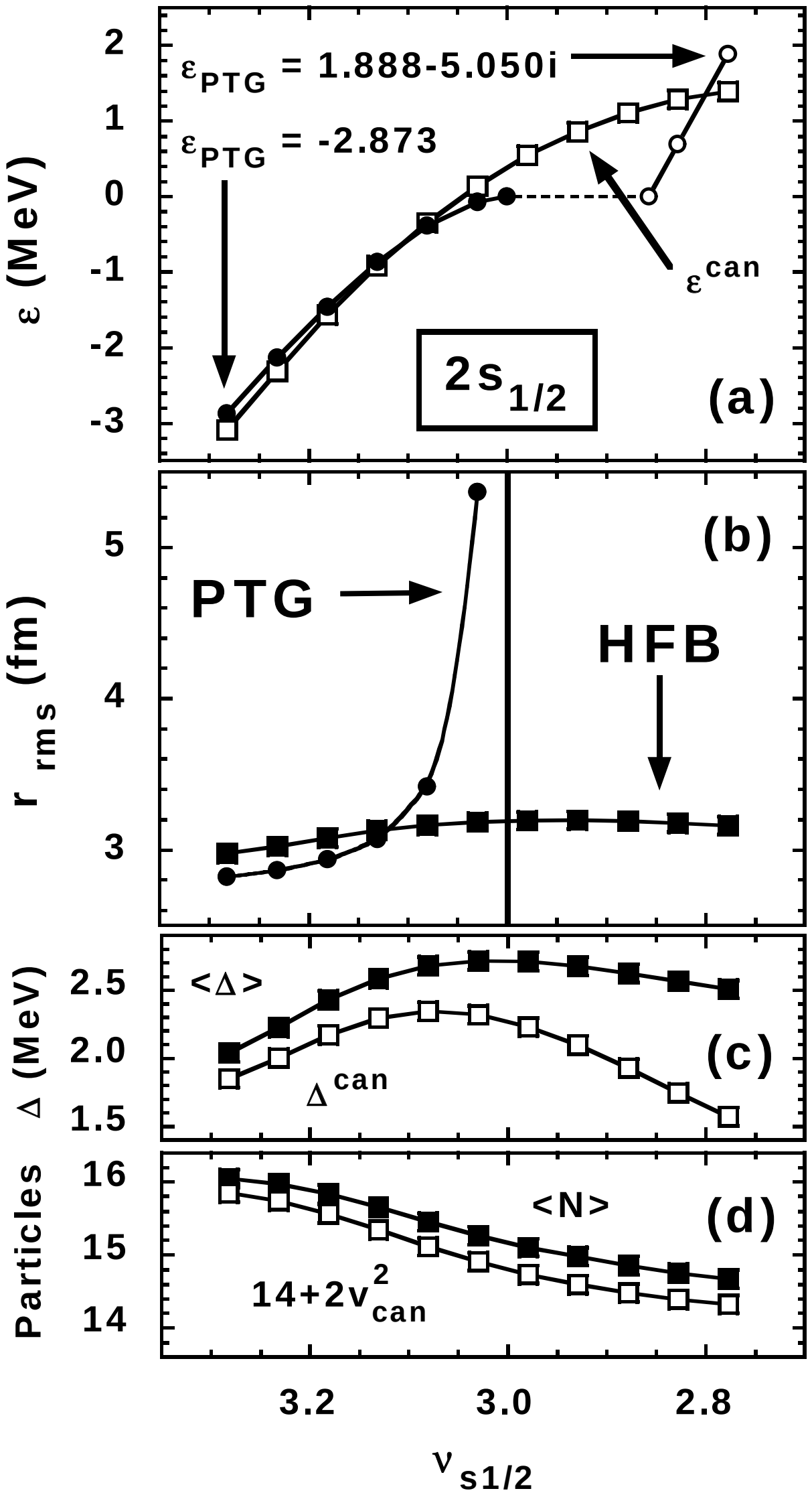

FIG. 2. Results of the spherical HFB calculations performed for the model PTG s.p. spectrum with varying positions of the $2s_{1/2}$ level. Panel (a) shows the HFB canonical s.p. energies (open squares), the PTG s.p. energies of the bound $2s_{1/2}$ states (full circles) and the real parts of the $2s_{1/2}$ resonances (open circles). The HFB neutron rms radius (b), average pairing gap $\langle\Delta\rangle$ (c), and average neutron number $\langle N\rangle$ (d) (full squares in panels (b)-(d)), were calculated for a fixed value of the neutron Fermi energy of $\lambda$=$-$0.056 MeV. The corresponding canonical pairing gap (c) and average neutron number (d) are shown with open squares. The no-pairing PTG rms radius in panel (b) corresponds to $N$=16.

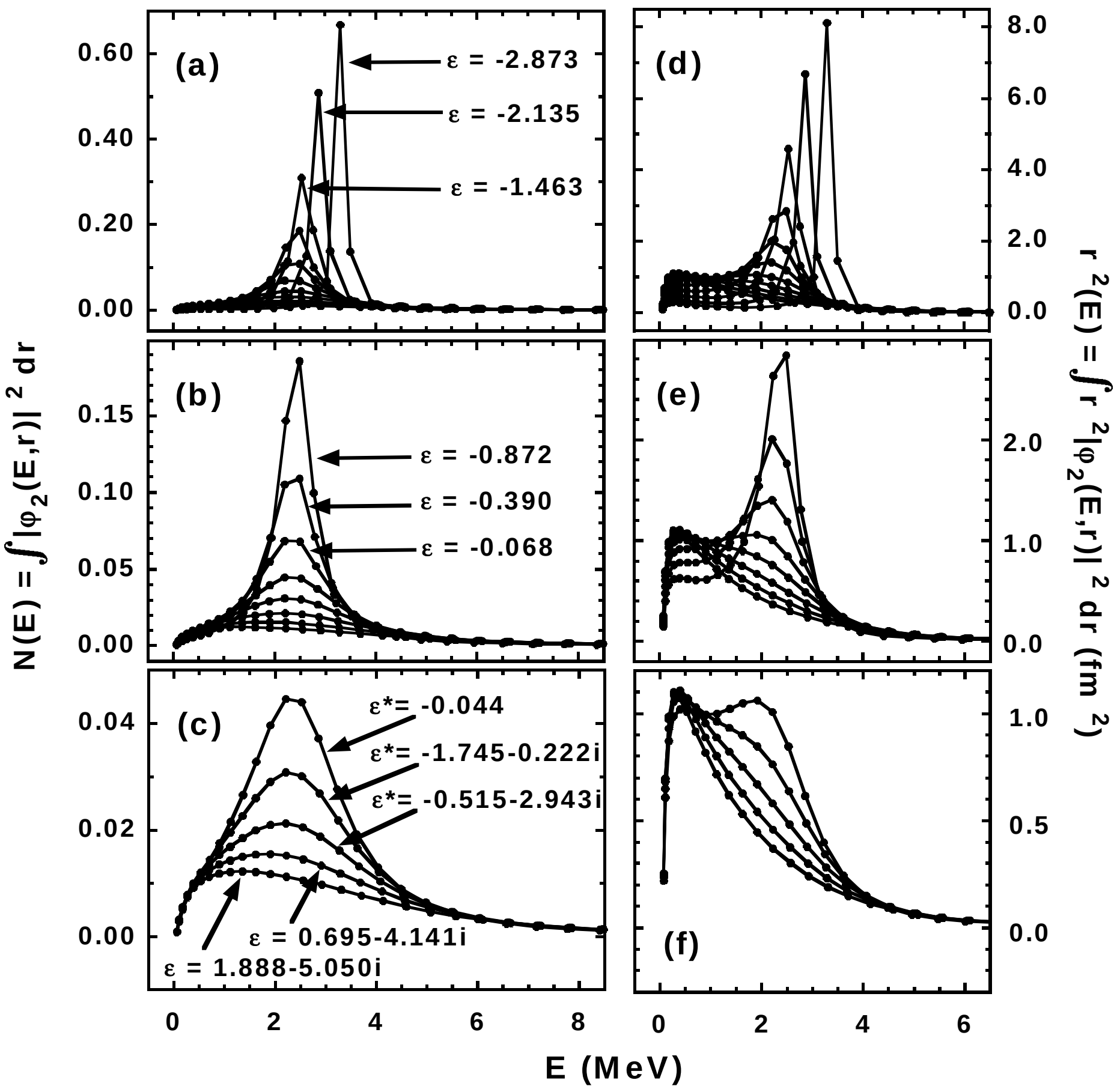

FIG. 3. Contributions $N(E)$ and $r^2(E)$ to the average neutron number and radius squared, respectively, calculated for the HFB q.p. states, as functions of the q.p. energy $E$. The HFB calculations have been performed at a fixed Fermi energy of $\lambda=-0.056$ MeV. The curves correspond to calculations performed for different positions of the $2s_{1/2}$ PTG states, with the s.p. energies indicated explicitly [see also Fig. 2(a)].